\documentclass[twocolumn,10pt, final, cleanfoot]{asme2ej}

\usepackage{graphicx} 
\usepackage{bm}
\usepackage{dcolumn}
\usepackage{amsmath}
\usepackage{textcomp}
\title{Deep Learning Convective Flow Using\\Conditional Generative Adversarial Networks}

\author{Changlin Jiang
    \affiliation{
	Mechanical and AI Lab (MAIL)\\
	Department of Mechanical Engineering\\
	Carnegie Mellon University\\
	Pittsburgh, Pennsylvania 15213\\
    Email: changlij@andrew.cmu.edu
    }	
}

\author{Amir Barati Farimani
    \affiliation{ 
    Assistant Professor \\
    Mechanical and AI Lab (MAIL)\\
    Department of Mechanical Engineering\\
    Carnegie Mellon University\\
	Pittsburgh, Pennsylvania 15213\\
    Email: barati@cmu.edu
    }
}

\begin{document}

\maketitle    

\begin{abstract}
{\it We developed a general deep learning framework, FluidGAN, capable of learning and predicting time-dependent convective flow coupled with energy transport. FluidGAN is thoroughly data-driven with high speed and accuracy and satisfies the physics of fluid without any prior knowledge of underlying fluid and energy transport physics. FluidGAN also learns the coupling between velocity, pressure, and temperature fields. Our framework helps understand deterministic multiphysics phenomena where the underlying physical model is complex or unknown. 
}
\end{abstract}




Convective transport is one of the most fundamental physical phenomena in fluid and transport, which can be applied to atmospheric circulation, nano-fluidic, micro-electrical systems, climates, and oceanography. With the emergence of massive data originating from weather stations and satellites, learning the physics of transport becomes amenable given a robust learning algorithm. Learning directly from transport data and the prediction based on the inference model is more accurate since the underlying noise in the data is learned. Energy transport can be described as the coupling between fluid momentum and highly non-linear energy. Computational Fluid Dynamics (CFD) has been used extensively to simulate and solve transport problems\cite{rappmicrofluidics}; however, CFD applications' time and memory consumption are usually prohibitively large\cite{brunton2016discovering}.

In recent years, deep learning has been making significant progress in almost all fields\cite{alom2018history}. Unlike traditional machine learning, which uses handmade features by a series of feature extraction algorithms, in the case of deep learning, the features are learned automatically and are represented hierarchically in multiple levels through analyzing massive datasets. In physics and engineering community, deep learning has introduced transformative solutions across diverse scientific disciplines\cite{brunton2016discovering, schaeffer2017learning, rudy2017data, mei2017mechanics, raissi2017physics, shen2023fishrecgan, fu2021transformer,  Mosser_2017, tompson2016accelerating,teng2021sketch2vis, Xie_2018, farimani2017deep}. However, most works are usually task-specific and still rely on understanding underlying physical rules. In this paper, we propose a FluidGAN model capable of inferring underlying physics and could directly predict stationary and time-dependent multi-physical phenomena using certain boundary conditions and initial conditions with both high accuracy and high speed, given sufficient computational or experimental data. Neither Navier-Stokes nor energy transport equations were given to the model. By using the conditional Generative Adversarial Network (cGAN)\cite{goodfellow2014generative, radford2015unsupervised, isola2017image}, a generator $G$, which trained adversarially with a discriminator $D$, could capture the distribution from multiphysics training data and make prediction $G(x, z)$ directly using condition $x$ and random noise $z$. Our results show that, although not required to do, our FluidGAN model could learn the pressure-momentum and energy-momentum couplings in convective transport. This generic FluidGAN model could be applied to many physical problems, both computationally and experimentally, in complex system prediction and physical law discovery. 

\begin{figure*}
\centering
\includegraphics{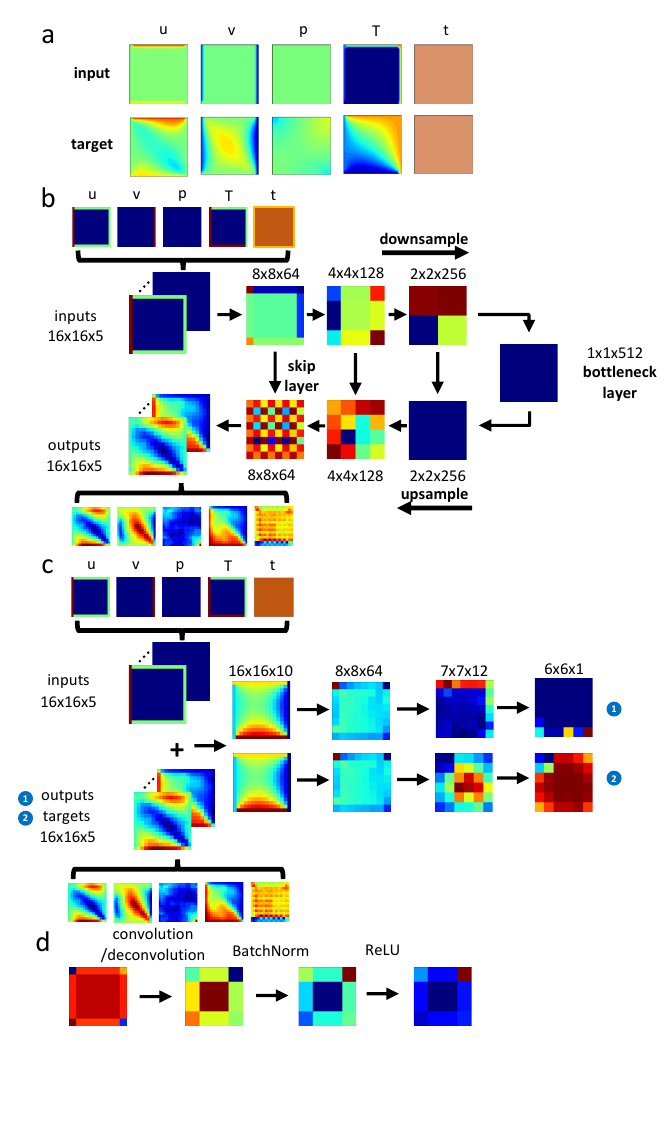}
\caption{\label{fig:Arc} \textbf{FluidGAN data representation and architecture}. This figure shows the time-dependent dataset setting. Stationary datset uses similar setting by skipping the time channel. \textbf{a}. Fluid data representation. The input represent boundary conditions (boundary grids) and initial conditions (interior grids). The target is the ground truth. Input and target have same time channel. \textbf{b}. The architecture of generator. Each arrow represents one layer described in \textbf{d}. \textbf{c}. The architecture of discriminator. Each arrow represents one layer described in \textbf{d}. The inputs will be concatenated by either outputs or targets. For outputs case (the first line), the discriminator score should be pretty low, represented by blue spots. In the contrary, targets case (the second line) should have high discriminator score, represented by red spots. \textbf{d}. The detailed architecture for each layer. Each layer consists of one convolution or deconvolution layer, one batchnorm and one ReLU activation.}
\end{figure*}
\begin{figure*}
\includegraphics{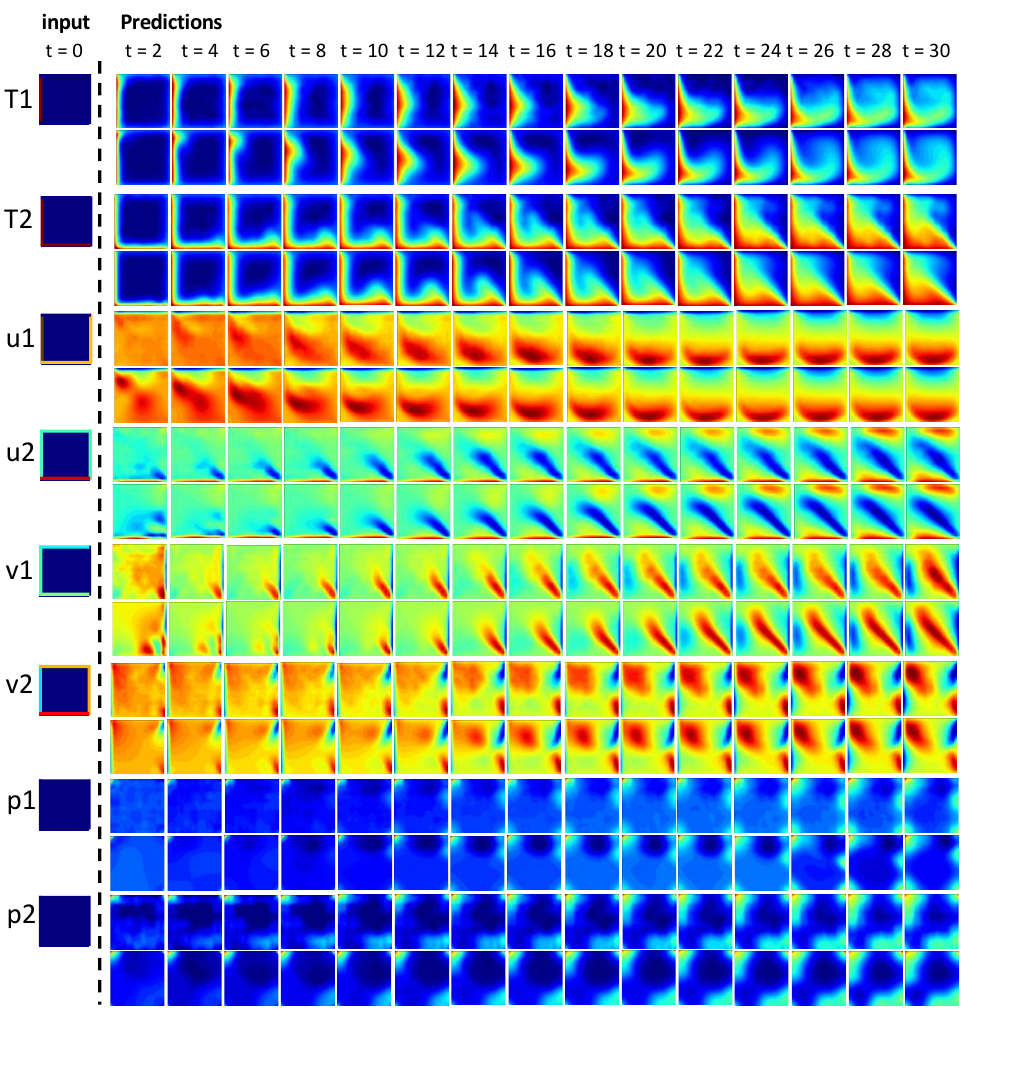}
\caption{\label{fig:TD_good} \textbf{Time-dependent convective flow prediction} This figure shows some good prediction results in the time-dependent dataset of $u$, $v$, $p$, $T$ respectively (in second). For each category, the first row is the prediction of FluidGAN model and the second row is the ground truth.}
\end{figure*}

We demonstrate this by studying a "benchmark" cavity convective transport problem in a square domain. According to the fluid dynamics rules, for the specific fluid variable $\Phi_{x, y}^{t}$, where $(x, y)\in M$ and $t\in T$, given its boundary conditions (BC) $\Phi_{(x,y)_{0}}$ and initial conditions (IC) $\Phi^{t_{0}}$, the value of $\Phi_{x, y}^{t}$ is deterministic. In other words, in laminar flow settings, the fluid variable of each time stamp maps to the unique boundary and initial conditions. Traditionally,  $\Phi_{x, y}^{t}$ is simulated using governing PDEs and compatible numerical methods. However,  taking advantage of the power of deep learning networks, when trained on a sizeable convective transport dataset that includes adequate kinds of possible fluid patterns, the FluidGAN model could directly and efficiently predict fluid variables from BCs and ICs without any previous understanding of underlying physics. The whole workflow can be divided into training and testing stages. In the training stage, we use ground truth fluid data, which could be either generated using computational or experimental methods, and its corresponding BCs/ICs to train a FluidGAN model. In testing, we use this pre-trained model to map BCs and ICs input to their output directly. 

The cavity flow dataset was generated using CFD software COMSOL\cite{COMSOL}. Concretely, the fluid variables $\Phi$ investigated in this paper are velocity field $u$ and $v$, pressure field $p$, and temperature field $T$. More details about dataset generation can be found in SI. As illustrated in Fig.~\ref{fig:Arc}(a), each sample of ground truth data (target) is a 16 $\times$ 16 $\times$ 4 tensor for steady state dataset, and 16 $\times$ 16 $\times$ 5 tensor for the time-dependent dataset with additional time channel. Besides, we apply a generic approach to "encode" our BCs/ICs information as input. Here, we use the boundary grids to represent boundary values/gradients and interior grids to represent initial conditions. For example, in the case of velocity $u$, the boundary grids represent the constant boundary conditions. The internal grids are all zero since we set the zero initial value for $u$. It's worthwhile to discuss the time channel representation here. For target data, the value of each grid in the time channel is simply set to the value of its corresponding time stamp. However, for input data, we could not use the value zero; the model couldn't tell the difference between inputs. Instead, we use the same time value as its corresponding target, acting as an "index" for the input. Note that this data representation approach is generic, and we can transfer this model to other multiphysics domains by modifying the number of channels. 

We now discuss the details of FluidGAN model. The FluidGAN consists of two networks: generator $G$ and discriminator $D$. The generator $G$ is trained to produce output $G(x, z)$ that cannot be distinguished from ground truth, using input condition $x$ and random noise vector $z$. In the meantime, we introduce another discriminator network $D$, which has the opposite objective as $G$. Specifically, the $D$ is trained to classify $G(x, z)$ as "fake" and classify ground truth data, $y$, as "real". By training $G$ and $D$ simultaneously and properly, the system will reach an equilibrium state where $G$ can make reasonable predictions, and $D$ is good at classifying unnatural predictions. Therefore, in the testing process, we can use the generator to make good predictions. The loss function of FluidGAN also reflects the concept of "adversarial", where $G$ tries to minimize this loss against an adversarial $D$ that tries to maximize it. The L1 loss is also added to capture the low-frequency feature.

\begin{equation}
\begin{aligned}
\textit{L}_{cGAN}(G, D) = &E_{x, y}[\log{D(x, y)}] + \\&E_{x, z}[\log{1 - D(x, G(x, z)}]
\label{cGAN loss}
\end{aligned}
\end{equation}
\begin{equation}
\begin{aligned}
\textit{L}_{L1}(G) = E_{x, y, z}[|| y - G(x, z) ||_{1}] 
\label{L1loss}
\end{aligned}
\end{equation}
\begin{equation}
\begin{aligned}
G^{*} = \arg \min \limits_G \max \limits_D \textit{L}_{cGAN}(G, D) + \lambda \textit{L}_{L1}(G)
\label{total objective}
\end{aligned}
\end{equation}

The architecture of generator $G$ and discriminator $D$ can be any network. Here, We adapt the architecture from cGAN, which uses an encoder-decoder generator and a PatchGAN discriminator as shown in  Fig.~\ref{fig:Arc}(b)(c). Each layer in both the generator and discriminator uses a module of the form convolution-BatchNorm-ReLU\cite{ioffe2015batch} in Fig.~\ref{fig:Arc}(d). The generator is a U-net network with a shipped connection layer. The encoder would extract hidden features to a bottleneck layer using a down-sampling process, while the decoder would sample the bottleneck layer to an output with the same shape as the input. This architecture ensures that the input and output can share much low-level information. To minimize possible information loss, we apply the skipped connection to this U-net and concatenate mirrored layers in the up-sampling process. The PatchGAN discriminator is actually a CNN network, which will produce an output layer. Each pixel of this output layer will represent a small "patch" in the original input. The idea behind PatchGAN is that it only restricts the attention to the structure in local input patches, and classifies if each patch is real or fake, thus achieving higher accuracy. As shown in Fig.~\ref{fig:Arc}(c), during one optimization step, both the output $G(x, z)$ and target $y$ are fed into the discriminator. The discriminator is then trained to predict a low score for $G(x,z)$ and a high score for $y$.
\begin{figure}[b]
\includegraphics{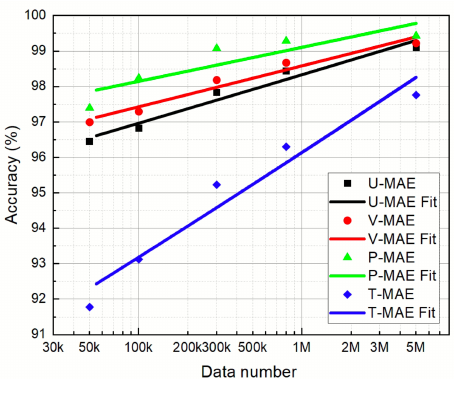}
\caption{\label{fig:MAE} \textbf{MAE for different number of training data, evaluated on time-dependent dataset}. The total number of time-dependent data is about 30M. We randomly took out a portion(50k, 100k, 300k, 800k and 5M) of data and split it into training set and testing set based on 80/20 ratio. We got over $99\%$ accuracy for the 5M dataset. }
\end{figure}

\begin{figure}[ht]
\includegraphics{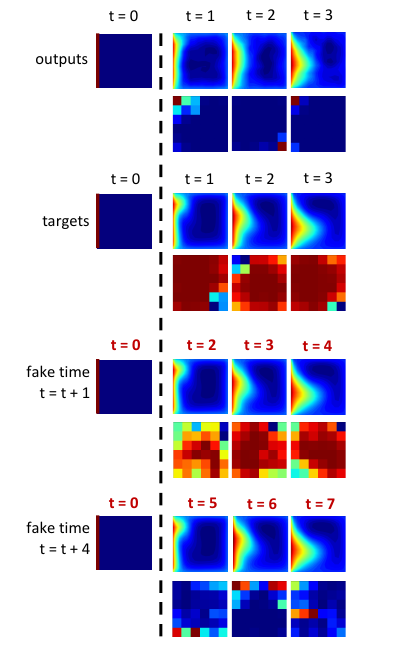}
\caption{\label{fig:Discri} \textbf{Discriminator performance for "fake time" inputs}. The first two rows show the output of discriminator for outputs and targets respectively, suggesting that the discriminator could tell the difference between outputs and targets. The last two rows show the output of discriminator for "fake time" inputs, in which we add t0 to the time channel. When t0 is getting bigger (i.e. from 0s to 4s), the discriminator could gradually classify this time channel dis-match.  }
\end{figure}

We use similar hyper-parameters as \cite{isola2017image} during the training process, as illustrated in supplementary.  The prediction results for both stationary and time-dependent datasets are shown in Fig.~\ref{fig:TD_good}, Fig.~\ref{fig:MAE}, and figures in the supplementary information.  These predicted videos perfectly capture the static states and intermediate fluid propagation states.  Also, the average residual continuity of prediction is $1.151\times10^{-3}$, close to the residual continuity of ground truth data $1.010\times10^{-3}$.  These results suggest high accuracy of the FluidGAN model.  Besides, since our model utilizes mainly convolution operation, it successfully learns the coupling between velocity, pressure, and temperature.  Recall the input pressure channel is filled with zero values representing the zero-gradient BC and IC, our model can't infer the pressure field solely from pressure channel input.  Our accurate pressure prediction demonstrates that our model learns the velocity-pressure coupling in convective transport through convolution-based networks.  We also get similar findings for velocity-temperature coupling, that is, inputs with the same temperature but different $u$, $v$ velocities map to different outputs.  Nevertheless, the additional variance of temperature input and the fact that temperature field and velocity/pressure fields are weakly coupled increase the difficulty of temperature prediction.  As demonstrated by Fig. \ref{fig:MAE}, the accuracy of the temperature channel is a little bit lower than the other three channels.  Luckily, increasing the number of training data mitigates this difficulty and boosts model generalizability. 

Another interesting finding is the performance of the discriminator.  Although the discriminator only affects the training process, it also exhibits excellent accuracy in discriminating generated results, or "fake" results, from ground truth results.  As Fig.~\ref{fig:Arc}(c) shows, although our generated outputs look pretty similar to the ground truth, the discriminator scores for prediction are overall very low (blue).  This result suggests that the FluidGAN discriminator is very robust in identifying nonphysical predictions.  We further explore this finding by constructing some "fake" predictions.  In the time-dependent prediction, we know that if we feed predictions into the discriminator, it will have a low discriminator score.  In the meantime, if we input targets to the discriminator, it should predict a high discriminator score, which should be close to 1.  To fool our model, we add some positive value $t0$ to the time channel of targets, namely combining  $u$, $v$, $p$, $T$ values of time $t$ with time value $t+t0$.  As Fig.~\ref{fig:Discri} suggests, when we increase $t0$, the discriminator can gradually tell the mismatch between the fake time channel and the actual time channel.  Therefore, the discriminator we get from the training could be used to identify nonphysical predictions, which may be further used to determine the quality of a numerical simulation. 

\begin{acknowledgment}
This work is supported by the start-up fund provided by CMU Mechanical Engineering. The authors would like to thank Zhonglin Cao, Nina Prakash, and Pranshu Pant for valuable comments and edits.
\end{acknowledgment}

%
% ...

\bibliographystyle{asmems4}

\bibliography{asme2e}

\begin{thebibliography}{10}

\bibitem{rappmicrofluidics}
Rapp, B.~E.
\newblock ``Microfluidics: Modeling, mechanics and mathematics''.

\bibitem{brunton2016discovering}
Brunton, S.~L., Proctor, J.~L., and Kutz, J.~N., 2016.
\newblock ``Discovering governing equations from data by sparse identification
  of nonlinear dynamical systems''.
\newblock {\em Proceedings of the National Academy of Sciences, {\bf 113}}(15),
  pp.~3932--3937.

\bibitem{alom2018history}
Alom, M.~Z., Taha, T.~M., Yakopcic, C., Westberg, S., Sidike, P., Nasrin,
  M.~S., Van~Esesn, B.~C., Awwal, A. A.~S., and Asari, V.~K., 2018.
\newblock ``The history began from alexnet: a comprehensive survey on deep
  learning approaches''.
\newblock {\em arXiv preprint arXiv:1803.01164}.

\bibitem{schaeffer2017learning}
Schaeffer, H., 2017.
\newblock ``Learning partial differential equations via data discovery and
  sparse optimization''.
\newblock {\em Proceedings of the Royal Society A: Mathematical, Physical and
  Engineering Sciences, {\bf 473}}(2197), p.~20160446.

\bibitem{rudy2017data}
Rudy, S.~H., Brunton, S.~L., Proctor, J.~L., and Kutz, J.~N., 2017.
\newblock ``Data-driven discovery of partial differential equations''.
\newblock {\em Science Advances, {\bf 3}}(4), p.~e1602614.

\bibitem{mei2017mechanics}
Mei, Y., Wang, S., Shen, X., Rabke, S., and Goenezen, S., 2017.
\newblock ``Mechanics based tomography: a preliminary feasibility study''.
\newblock {\em Sensors, {\bf 17}}(5), p.~1075.

\bibitem{raissi2017physics}
Raissi, M., Perdikaris, P., and Karniadakis, G.~E., 2017.
\newblock ``Physics informed deep learning (part ii): Data-driven discovery of
  nonlinear partial differential equations''.
\newblock {\em arXiv preprint arXiv:1711.10566}.

\bibitem{shen2023fishrecgan}
Shen, X., Joo, K., and Oh, J., 2023.
\newblock ``Fishrecgan: An end to end gan based network for fisheye
  rectification and calibration''.
\newblock {\em arXiv preprint arXiv:2305.05222}.

\bibitem{fu2021transformer}
Fu, Q., Teng, Z., White, J., and Schmidt, D.~C., 2021.
\newblock ``A transformer-based approach for translating natural language to
  bash commands''.
\newblock In 2021 20th IEEE International Conference on Machine Learning and
  Applications (ICMLA), IEEE, pp.~1245--1248.

\bibitem{Mosser_2017}
Mosser, L., Dubrule, O., and Blunt, M.~J., 2017.
\newblock ``Reconstruction of three-dimensional porous media using generative
  adversarial neural networks''.
\newblock {\em Physical Review E, {\bf 96}}(4), Oct.

\bibitem{tompson2016accelerating}
Tompson, J., Schlachter, K., Sprechmann, P., and Perlin, K., 2016.
\newblock Accelerating eulerian fluid simulation with convolutional networks.

\bibitem{teng2021sketch2vis}
Teng, Z., Fu, Q., White, J., and Schmidt, D.~C., 2021.
\newblock ``Sketch2vis: Generating data visualizations from hand-drawn sketches
  with deep learning''.
\newblock In 2021 20th IEEE International Conference on Machine Learning and
  Applications (ICMLA), IEEE, pp.~853--858.

\bibitem{Xie_2018}
Xie, Y., Franz, E., Chu, M., and Thuerey, N., 2018.
\newblock ``tempogan''.
\newblock {\em ACM Transactions on Graphics, {\bf 37}}(4), Jul, p.~1–15.

\bibitem{farimani2017deep}
Farimani, A.~B., Gomes, J., and Pande, V.~S., 2017.
\newblock Deep learning the physics of transport phenomena.

\bibitem{goodfellow2014generative}
Goodfellow, I., Pouget-Abadie, J., Mirza, M., Xu, B., Warde-Farley, D., Ozair,
  S., Courville, A., and Bengio, Y., 2014.
\newblock ``Generative adversarial nets''.
\newblock In Advances in neural information processing systems, pp.~2672--2680.

\bibitem{radford2015unsupervised}
Radford, A., Metz, L., and Chintala, S., 2015.
\newblock ``Unsupervised representation learning with deep convolutional
  generative adversarial networks''.
\newblock {\em arXiv preprint arXiv:1511.06434}.

\bibitem{isola2017image}
Isola, P., Zhu, J.-Y., Zhou, T., and Efros, A.~A., 2017.
\newblock ``Image-to-image translation with conditional adversarial networks''.
\newblock In Proceedings of the IEEE conference on computer vision and pattern
  recognition, pp.~1125--1134.

\bibitem{COMSOL}
{COMSOL Multiphysics}.
\newblock Comsol v5.3a.

\bibitem{ioffe2015batch}
Ioffe, S., and Szegedy, C., 2015.
\newblock Batch normalization: Accelerating deep network training by reducing
  internal covariate shift.

\end{thebibliography}

\appendix       
\section*{Appendix A: Dataset detail} To simplify, we assume laminar, incompressible Newtonian fluid with no-slip boundary condition. Also, we assume the flow has constant density $\rho$, viscosity $\nu$ and no source term. We set Dirichlet boundary conditions for $u$, $v$ and $T$, and Neumann boundary condition for pressure with gradient zero. Zero initial value of $u$, $v$, $p$, $T$ are also assumed. During data generation, the velocity is ranging from -2 $m/s$ to 2 $m/s$ with step of 0.5$m/s$, and temperature is ranging from 0\textdegree{}C to 20\textdegree{}C with step of 5\textdegree{}C. In setting we generated 65000 different steady state fluid data combinations. For time dependent dataset, we generated 50000 BC/IC videos, each  consisting  of  600 frames. The time step between each frame is 0.1s and the total simulation time is 60s. Then whole data are then randomly split into train/test sets following an 80/20 ratio.

\section*{Appendix B: Training detail}
During training process, we use similar hyper-parameters as \cite{isola2017image}. In specific, we use batch size of 1, learning rate $\alpha_{G} = 2\times10^{-4}$ for generator and $\alpha_{D} = 2\times10^{-5}$ for discriminator. The ratio of GAN loss and L1 loss is 1:1000. The optimizer we use is the commonly used Adam optimizer with $\beta_{1} = 0.5$.  For the training and testing we use Nvidia GeForce GTX 1080 Ti GPU with 11GB RAM and Intel® Core™ i7-8700K CPUs. Also, we used the tensorflow v1.12 deep learning framework. Code and dataset is available in \verb+https://github.com/BaratiLab/FluidGAN-TensorFlow+

\begin{figure*}
\includegraphics{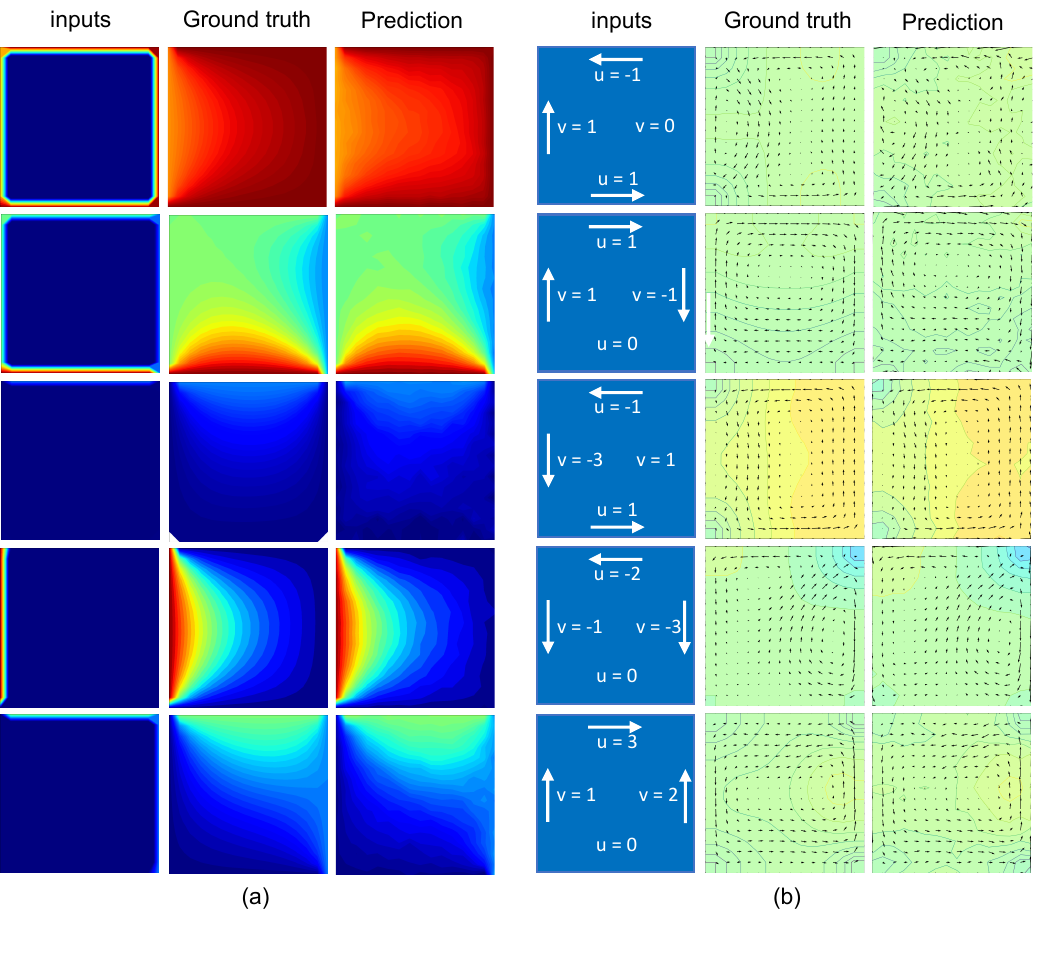}
\caption{\label{fig:ST_all} \textbf{Prediction for stationary dataset} \textbf{a.} Prediction results of temperature channel. The boundary nodes of inputs encode the Dirichlet temperature boundary condition, and the interior nodes of inputs represent the inital values, which are zero in our case. \textbf{b.} Prediction results of velocity and pressure channels. The inputs encode BC/IC of velocity and pressure. Arrows in targets and outputs represent velocity vector, and the background contour represents pressure field. }
\end{figure*}

\begin{figure*}
\includegraphics{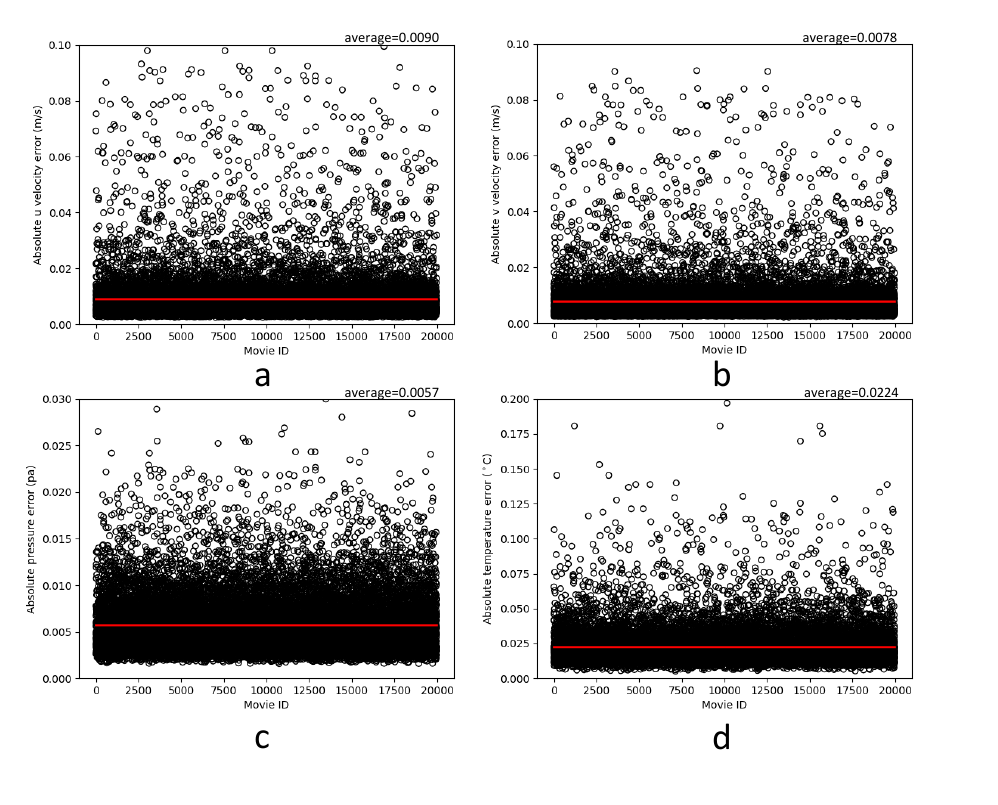}
\caption{\label{Error_distribution} \textbf{Statistical accuracy of FluidGAN for $u$, $v$, $p$ and $T$, evaluated on test time-dependent dataset } \textbf{a}. MAE of u velocity \textbf{b}. MAE of v velocity \textbf{c}. MAE of pressure \textbf{d}. MAE of temperature. The red line is the average value of each MAE. The results are all evaluated without normalization.}
\end{figure*}

\begin{figure*}
\includegraphics{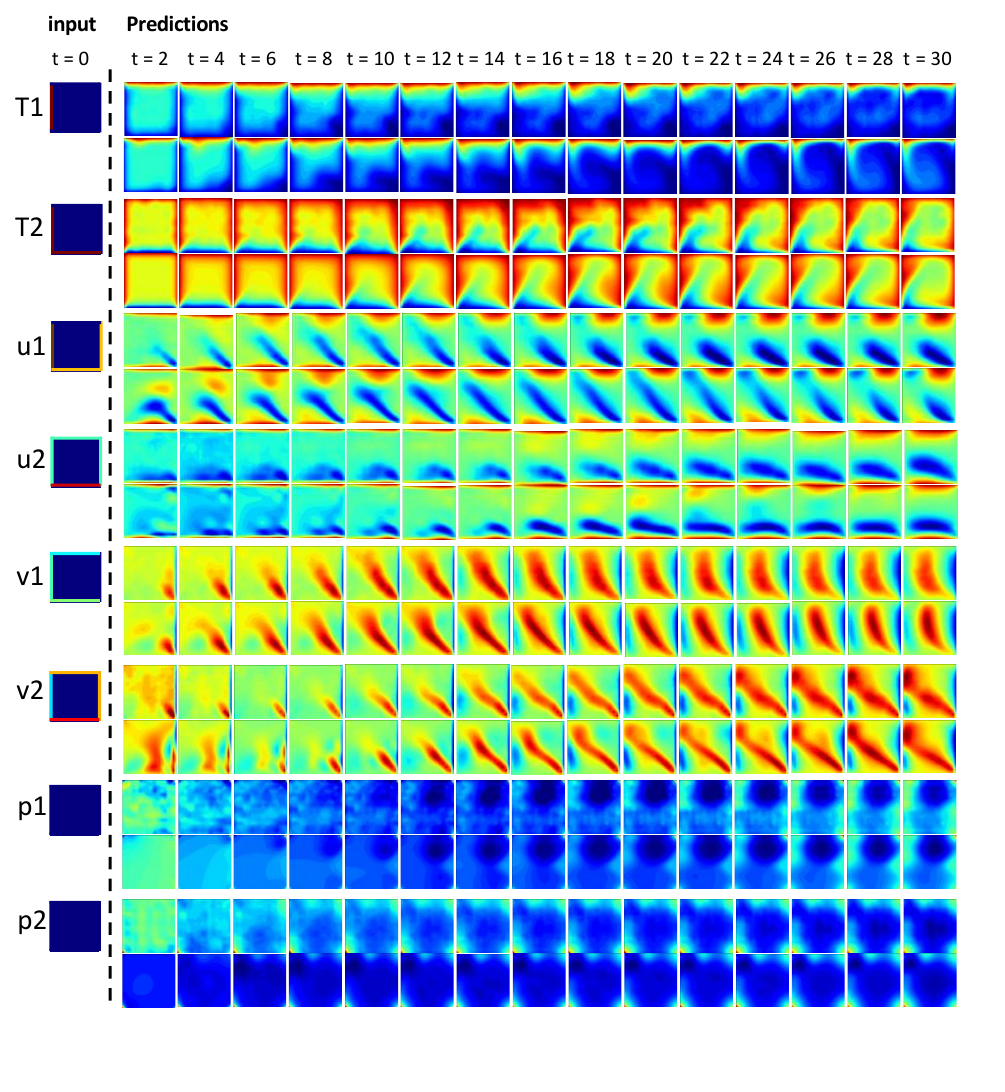}
\caption{\label{TD_bad} \textbf{Some less satisfactory results of time-dependent dataset} For each category, the first row is the predictions of FluidGAN and the second row is the ground truth. The model sometimes fail to keep track of the time channel, that is, to predict beforehand or delayed results.}
\end{figure*}

\end{document}